\documentclass[12pt]{article}
\usepackage{epsfig}
\usepackage{a4wide}

\newcommand{\pp}{\phantom{-}}

\newcommand{\lam}{\lambda}
\newcommand{\Lam}{\Lambda}

\newcommand{\pf}{\protect \boldmath}
\newcommand{\lbf}{\Large\bf}

\begin{document}
\thispagestyle{empty}

\begin{flushright}
        MZ-TH/03-06\\
        hep-ph/0309047\\
        September 2003\\
\end{flushright}

\vspace{5mm}

\begin{center}
 {\lbf Semi-inclusive decays \pf$\Lam_b\rightarrow X_c+(D_s, D_s^*)$ at
  $O(\alpha_s)$\\[3mm]
  including \pf$\Lam_b$ and $D_s^*$ polarization effects}\\[7mm]
 {\large M. Fischer, S.~Groote, J.G.~K\"orner and M.C.~Mauser}\\[7mm]
       Institut f\"ur Physik, Johannes-Gutenberg-Universit\"at\\[2mm]
       Staudinger Weg 7, D--55099 Mainz, Germany\\[7mm]
\end{center}

\vspace{7mm}


\begin{abstract}\noindent
In the leading order of the $(1/m_b)$-expansion in HQET the dominant
contribution to the semi-inclusive decays of polarized $\Lam_b$ baryons into
the charm--strangeness mesons $D_s$ and $D_s^*$ is given by the partonic
process $b(\uparrow)\rightarrow c+(D_s^-,D_s^{*-})$. Using standard values for
the parameters of the process one expects a rather large branching ratio of
$\approx 8\%$ into these two channels. In the factorization approximation the
semi-inclusive decay of a polarized $\Lam_b$ is governed by three unpolarized
and four polarized structure functions for which we determine the
nonperturbative $O(1/m_b^2)$ corrections and the $O(\alpha_s)$ radiative
corrections. We find that the perturbative and nonperturbative corrections
amount to $\approx 10\%$ and $\approx 3\%$, respectively. The seven structure
functions can be measured through an analysis of the joint decay distributions
of the process involving the polarization of the $\Lam_b$ and the decays
$D_s^{*-}\rightarrow D_s^-+\gamma$ and $D_s^{*-}\rightarrow D_s^-+\pi^0$ for
which we provide explicit forms. We also provide numerical results for the
Cabibbo suppressed semi-inclusive decays $\Lam_b\rightarrow X_u+(D_s,D_s^*)$.
\end{abstract}

\newpage


\section{\bf Introduction} 
 
In the leading order of the $(1/m_b)$-expansion in HQET the semi-inclusive
decay $\Lam_b\rightarrow X_c+(D_s^-,D_s^{*-})$ is dominated by the partonic
process $b\rightarrow c+(D_s^-,D_s^{*-})$. The basic assumption is that
factorization holds for the nonleptonic decay process
$\Lam_b\rightarrow X_c+(D_s^-,D_s^{*-})$. One can then factorize the
semi-inclusive decay into a current-induced $\Lam_b\rightarrow X_c$ transition
and a current-induced vacuum one-meson transition. The leading order $1/m_Q$
contribution to the $\Lam_b\rightarrow X_c$ transition is given by the
partonic $ b\rightarrow c$ transition. There are two types of corrections to
the leading order result. First there are the nonperturbative corrections
which set in at $O(1/m_b^2)$ in the heavy mass expansion. They can be
estimated using the methods of the operator product expansion in HQET. Second
there are also the perturbative $O(\alpha_s)$ corrections which can be
calculated using standard techniques. From a previous calculation of the
corresponding decays in the mesonic sector
$\bar{B^0}\rightarrow X_c+(D_s^-,D_s^{*-})$ one expects perturbative and
nonperturbative corrections of $\approx 10\%$ \cite{aleksan00,fgkm00} and
$\approx 1\%$ \cite{fgkm00}, respectively. 

When the $\Lam_b$ is unpolarized, the decay
$\Lam_b\rightarrow X_c+(D_s^-, D_s^{*-})$ is quite similar to the
corresponding mesonic decay $\bar{B^0}\rightarrow X_c+(D_s^-,D_s^{*-})$
\cite{aleksan00,fgkm00}. In fact, to leading order in the $1/m_b$ expansion
and to any order in the perturbative QCD corrections the two semi-inclusive
decays are identical to one another. However, when the $\Lam_b$ is polarized,
there are four additional polarized structure functions that enter the decay
analysis. One can thus probe four more structure functions in the
semi-inclusive decay of a polarized $\Lam_b$ than it is possible in the
corresponding $B$-meson decay. Polarized $b$-quarks and thereby polarized
$\Lam_b$ baryons arise quite naturally in weak decays such as
$Z\rightarrow b\bar b$ and $t\rightarrow Wb$. When the polarized $b$-quark
fragments into a $\Lam_b$ baryon, $\approx 70\%$ of its polarization is
retained \cite{ckps92,fp}.

We mention that large samples of $\Lam_b$'s are expected to be produced at
the currently running $p\bar p$ collider Tevatron 2. In fact the first few
$\Lam_b$'s have been reconstructed by the CDF collaboration using the
superior tracking capacity of their new silicon vertex trigger \cite{vila03}.


\section{\bf Angular decay distributions}

In the factorization approximation the semi-inclusive decays of polarized
$\Lam_b$ baryons\break $\Lam_b(\uparrow)\rightarrow X_c+(D_s^-,D_s^{*-})$ are
governed by altogether seven structure functions which can be measured by an
angular analysis of the decay process. We mention that there are two
additional parity-violating structure functions in the decays
$\Lam_b(\uparrow)\rightarrow X_c+D_s^{*-}$ which, however, cannot be measured
since the dominating decays of the $D_s^{*-}$ are parity-conserving. 
 
Five of the structure functions describe the semi-inclusive decay
$\Lam_b\rightarrow X_c+D_s^{*-}$ into vector mesons followed by their
subsequent decay into $D_s^{*-}\rightarrow D_s^-+\gamma$ and
$D_s^{*-}\rightarrow D_s^-+\pi^0$. The branching ratios of the $D_s^{*-}$ into
these two principal channels are given by $(94.2\pm 2.5)\%$ and
$(5.8\pm 2.5)\%$ \cite{pdg02}, respectively.

\begin{figure}[ht]\begin{center}
\epsfig{file=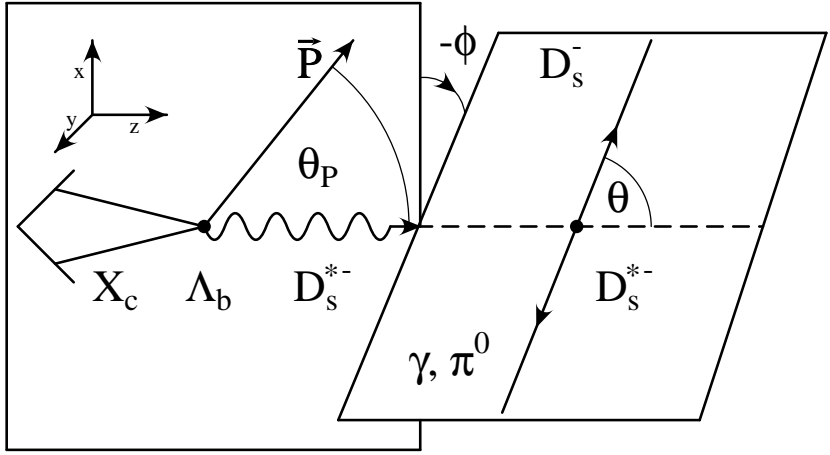,scale=0.6}
\caption{\label{fig1}Definition of polar angles $\theta_P$ and $\theta$, and
the azimuthal angle $\phi$ in the semi-inclusive decay $\Lam_b(\uparrow)$
$\rightarrow X_c+D_s^*$ $(\rightarrow D_s^-+\gamma\mbox{ or }\pi^0)$. $\vec P$
is the polarization vector of the $\Lam_b$. The polar angle $\theta$ is
defined in the $D_s^{*-}$ rest frame relative to the direction of the
$D_s^{*-}$ in the $\Lambda_b$ rest frame.}
\end{center}\end{figure}

The angular decay distribution of the semi-inclusive polarized $\Lam_b$ decays
can be obtained from the master formula (see e.g.\ \cite{fgkm02})
\begin{equation} 
\label{masterformula}
  W(\theta_P, \theta, \phi) \propto
  \hspace{-8mm} \sum\limits_{
  \lam_{D_s^*}-\lam^{\prime}_{D_s^*} =
  \lam_{\Lam_b}-\lam^{\prime}_{\Lam_b}}
  \hspace{-8mm} 
  e^{i (\lam_{D_s^*}-\lam^{\prime}_{D_s^*}) \phi} \,
  d^1_{\lam_{D_s^*} m} (\theta) \,
  d^1_{\lam^{\prime}_{D_s^*} m} (\theta) \,
  H_{\lam_{D_s^*} \lam^{\prime}_{D_s^*}}%
   ^{\lam_{\Lam_b} \: \lam^{\prime}_{\Lam_b}}\,
  \rho_{\lam_{\Lam_b} \: \lam^{\prime}_{\Lam_b}} (\theta_P),
\end{equation}
where
$ \rho_{\lam_{\Lam_b} \: \lam^{\prime}_{\Lam_b}} (\theta_P) $
is the density matrix of the $ \Lam_b $ which reads
\begin{equation} 
  \rho_{\lam_{\Lam_b} \: \lam^{\prime}_{\Lam_b}} (\theta_P) =
  \frac{1}{2} \pmatrix{
   1+P \cos \theta_P  &  P \sin \theta_P \cr
   P \sin \theta_P  &  1-P \cos \theta_P }.
\end{equation}
$P$ is the magnitude of the polarization of the $\Lam_b$. The
$H_{\lam_{D_s^*}\lam'_{D_s^*}}^{\lam_{\Lam_b}\:\lam'_{\Lam_b}}$ are the
helicity components of the hadronic tensor $H_{\mu\nu}(s_{\Lam_b})$ describing
the semi-inclusive decay. The sum in Eq.~(\ref{masterformula}) extends over
all values of $\lam_{D_s^*},\lam'_{D_s^*},\lam_{\Lam_b}$ and $\lam'_{\Lam_b}$
compatible with the constraint $\lam_W-\lam'_W=\lam_{\Lam_b}-\lam'_{\Lam_b}$
(the spin degrees of freedom of $X_c$ are being summed over). The polar angles
$\theta_P$, $\theta$ and the azimuthal angle $\phi$ are defined in
Fig.~\ref{fig1}.
Because of angular momentum conservation, the second lower index in the small
Wigner $d(\theta)$-function $d^1_{\lam_{D_s^*}m}(\theta)$ runs over $m =\pm 1$
for the decay $D_s^{*-}\rightarrow D_s^-+\gamma$ and over $m=0$ for the decay
$D_s^{*-}\rightarrow D_s^-+\pi^0$. One thus obtains the angular decay
distributions
\begin{eqnarray} 
\label{angular1} 
 \frac{d \Gamma_{\! \Lam_b^{(\uparrow)} \rightarrow X_c+D_s^{*-}
 (\rightarrow D_s^{-}+\gamma)}}
 {d\!\cos \theta_P\,d\!\cos \theta\,d\!\phi} & = &
 \frac{1}{4 \pi} \mbox{BR}(D_s^{*-} \!\rightarrow\! D_s^{-}+\gamma)
 \Bigg\{ \frac{3}{8} (1+\cos^2 \theta)
 (\Gamma_U+\Gamma_{U^P} P \cos \theta_P ) \nonumber \\[2mm] &+&
 \frac{3}{4} \sin^2 \theta
 (\Gamma_L+\Gamma_{L^P} P \cos \theta_P) +
 \frac{3}{4} \sqrt{2} P \sin \theta_P \sin 2 \theta \cos \phi \,
 \Gamma_{I^P} \!\! \Bigg\} \nonumber \\
\end{eqnarray}
and
\begin{eqnarray} 
 \label{angular2}
 \frac{d \Gamma_{\! \Lam_b^{(\uparrow)} \rightarrow X_c+D_s^{*-} 
 (\rightarrow D_s^{-}+\pi^0)}}
 {d\!\cos \theta_P\,d\!\cos \theta\,d\!\phi} & = &
 \frac{1}{4 \pi} \mbox{BR}(D_s^{*-} \!\rightarrow\! D_s^{-}+\pi^{0}) 
 \Bigg\{ \frac{3}{4} \sin^2 \theta
 (\Gamma_U+\Gamma_{U^P} P \cos \theta_P ) \nonumber \\[2mm] &+&
 \frac{3}{2} \cos^2 \theta
 (\Gamma_L \!+\! \Gamma_{L^P} P \cos \theta_P) \!-\!
 \frac{3}{2} \sqrt{2} P \sin \theta_P \sin 2 \theta \cos \phi \,
 \Gamma_{I^P} \!\! \Bigg\}. \nonumber \\
\end{eqnarray} 
Two-fold or single angle decay distributions can be obtained from
Eqs.~(\ref{angular1},\ref{angular2}) by further integration. For example, the
single angle dependence on $\cos\theta_P$ for both cases is given by
\begin{eqnarray} 
\label{singleangle-star}
\frac{d \Gamma_{\! \Lam_b^{(\uparrow)} \rightarrow X_c+D_s^{*-} }}
 {d\!\cos \theta_P} & = & \frac{1}{2}
 (\Gamma_{U+L}+\Gamma_{(U+L)^P} P \cos \theta_P ) \nonumber \\[2mm]
                    & = & \frac{1}{2} \Gamma_{U+L}
 (1+\alpha_P(D_s^*) P \cos \theta_P ),
\end{eqnarray}
where we have defined an asymmetry parameter 
$ \alpha_P(D_s^*) = \Gamma_{(U+L)^P} / \Gamma_{U+L} $.

The transverse/longitudinal composition of the vector meson $D_s^{*-}$ can be
best determined by analyzing the $\cos\theta $-dependence of the decay
distributions after integrating over $\cos\theta_P$ and $\phi$. Note that the
$\cos\theta$-dependence is different in the two decay modes.

The decay distribution for $\Lam_b(\uparrow)\rightarrow X_c+D_s^-$ can be
obtained from the same master formula (\ref{masterformula}) with the
appropiate substitutions $\lam_{D_s^*}\rightarrow\lam_{D_s}=0$ and
$d^1\rightarrow d^0=1$. The helicity of the $\lam_{D_s}$ will be denoted by
the symbol ``$S$'' for ``scalar''. One has 
\begin{eqnarray}
\label{singleangle}
 \frac{d \Gamma_{\Lam_b^{(\uparrow)} \rightarrow X_c+D_s^-}}
 {d \!\cos \theta_P} & = &
 \frac{1}{2} \left( \Gamma_S +
 \Gamma_{S^P} P \cos \theta_P \right) \nonumber \\[2mm] 
                     & = & \frac{1}{2} \Gamma_S
 (1+\alpha_P(D_s) P \cos \theta_P ) ,
\end{eqnarray}
where we have again defined an asymmetry parameter 
$\alpha_P(D_s)=\Gamma_S^P/\Gamma_S$.

The angular coefficients $\Gamma_i$ ($i= S, S^P, U, L, U^P, L^P, I^P$)
appearing in the decay distributions are partial helicity rates defined by
\begin{equation}
\label{partialrates}
 \Gamma_i  = \frac{G_F^2}{8 \pi}
 |V_{bc} V_{cs}^{\ast}|^2 f_{D_s^{(\ast)}}^2 m_b^2 \,
 p_{D_s^{(*)}}\,a_1^2\,H_i,
\end{equation}
where the helicity structure functions $H_i$ are linear combinations of the
helicity components. They read
\begin{eqnarray}
\label{system3anfang}
  H_{S^{\phantom P}} & = & \phantom{\frac{1}{4} (}
  H^{++}_{S\,S}+H^{--}_{S\,S}, \nonumber\\ 
  H_{U^{\phantom P}} & = & \phantom{\frac{1}{4} (}
  H_{++}^{++}+H_{++}^{--}+H_{--}^{++}+H_{--}^{--}, \nonumber\\
  H_{L^{\phantom P}} & = & \phantom{\frac{1}{4} (}
  H_{0 \: 0}^{++}+H_{0 \: 0}^{--}, \nonumber\\
  H_{S^P} & = & \phantom{\frac{1}{4} (}
  H^{++}_{S\,S}-H^{--}_{S\,S}, \\
  H_{U^P} & = & \phantom{\frac{1}{4} (}
  H_{++}^{++}-H_{++}^{--}+H_{--}^{++}-H_{--}^{--}, \nonumber\\
  H_{L^P} & = & \phantom{\frac{1}{4} (}
  H_{0 \: 0}^{++}-H_{0 \: 0}^{--}, \nonumber\\
  H_{I^P} & = & \frac{1}{4} (
  H_{+\,0}^{+-}+H_{0 \: +}^{-+} -
  H_{-\,0}^{-+}-H_{0 \: -}^{+-}) =
  \frac{1}{2} (H_{+\,0}^{+-}-H_{-\,0}^{-+}),\nonumber
\end{eqnarray}
where, for the ease of writing, we have omitted factors of $1/2$ in the upper
indices standing for the helicities of the $\Lam_b$. The remaining quantities
appearing in (\ref{partialrates}) are defined in Sec.~3.

When the $\Lam_b$ is unpolarized ($P=0$), or when one integrates over the
angles $\theta_P$ and $\phi$ that describe the orientation of the polarization
vector of the $\Lam_b$, one remains with the contributions of the three
structure functions $H_U$, $H_L$ and $H_S$ in the decay distributions. In this
way one recovers the decay distributions for the corresponding semi-inclusive
decays of $B$ mesons into $D_s$ and $D_s^*$ treated in \cite{fgkm00}.  


\section{\pf Born term rates}

\begin{figure}[ht]\centering\leavevmode
\put(0,60){\bf a)}\epsfig{file=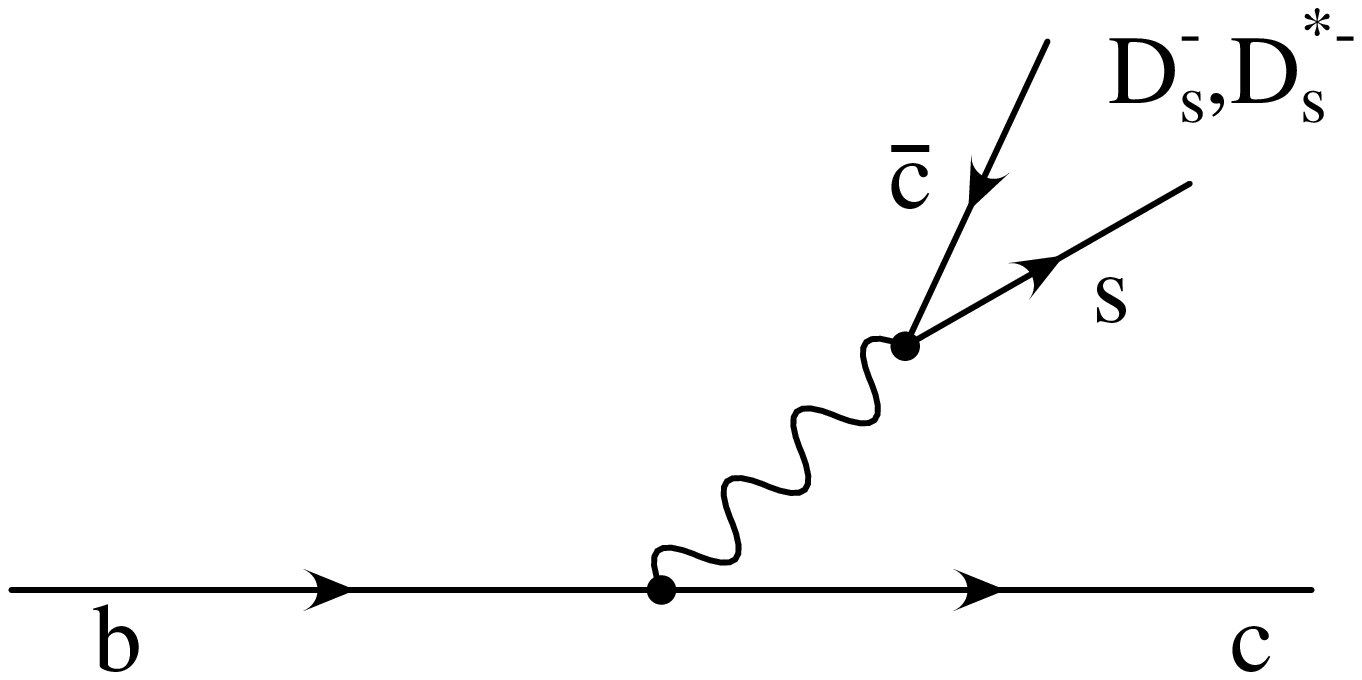,scale=0.4}\kern64pt
\put(0,60){\bf b)}\epsfig{file=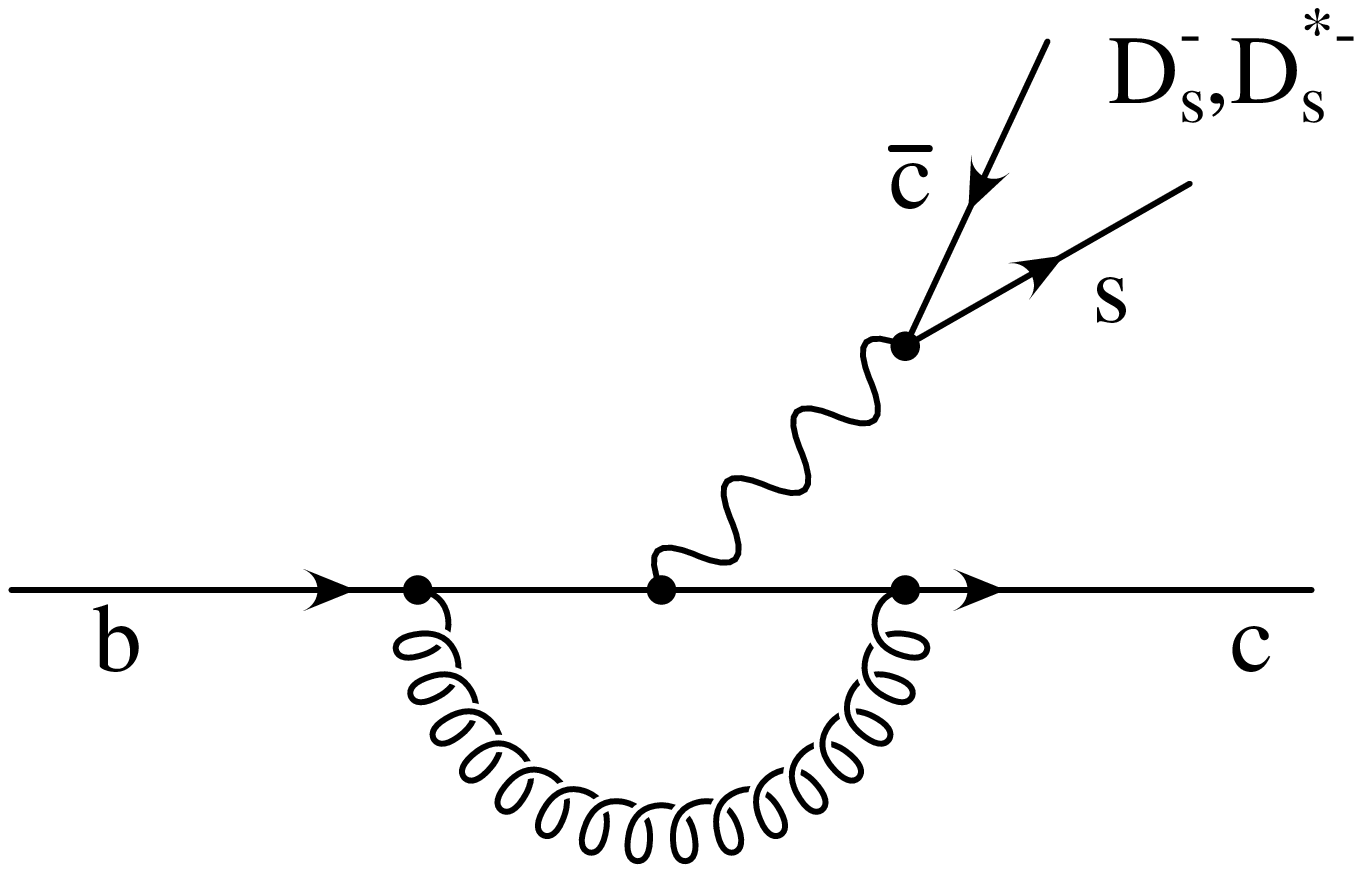,scale=0.4}\newline\noindent
\put(0,60){\bf c)}\epsfig{file=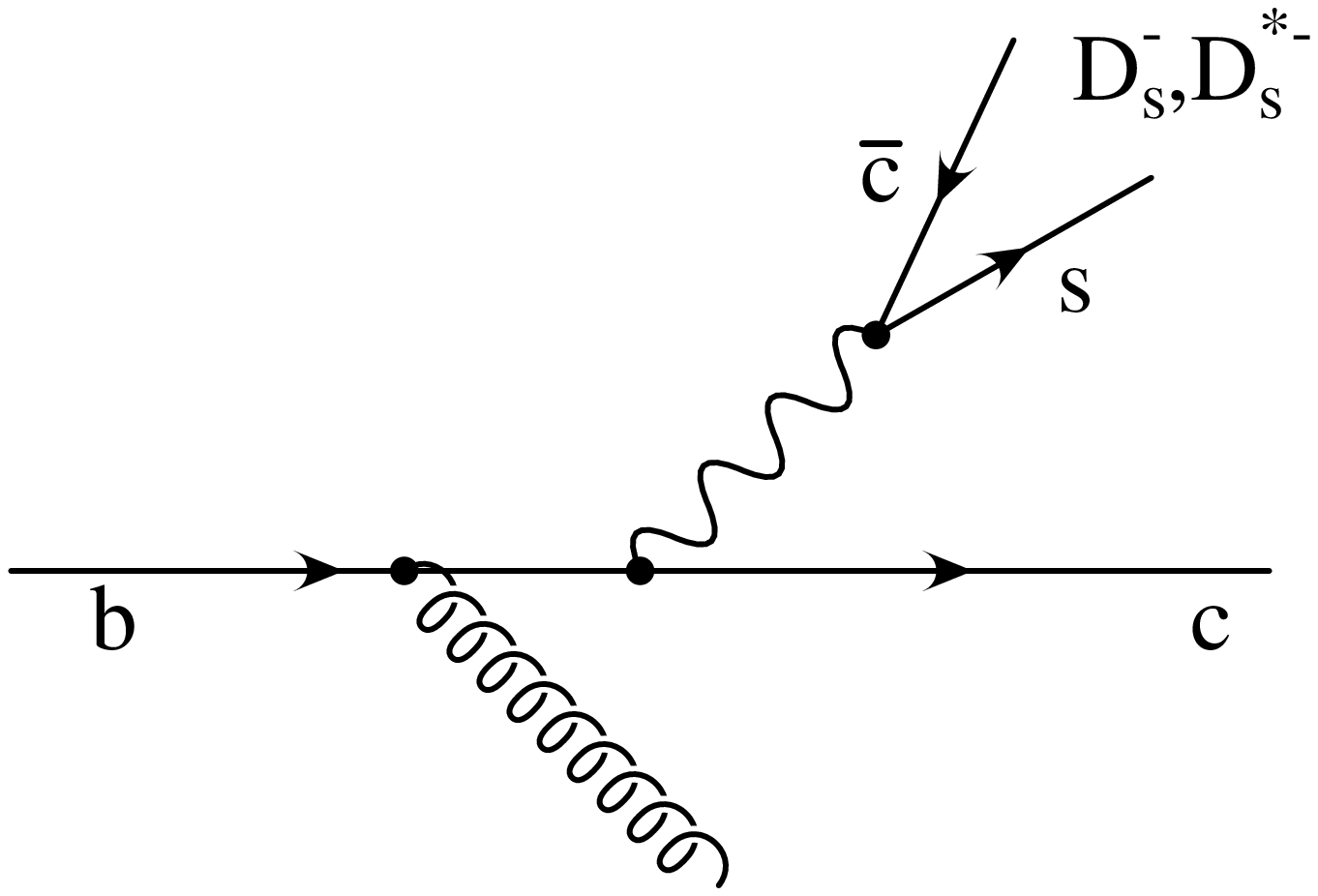,scale=0.4}\kern64pt
\put(0,60){\bf d)}\epsfig{file=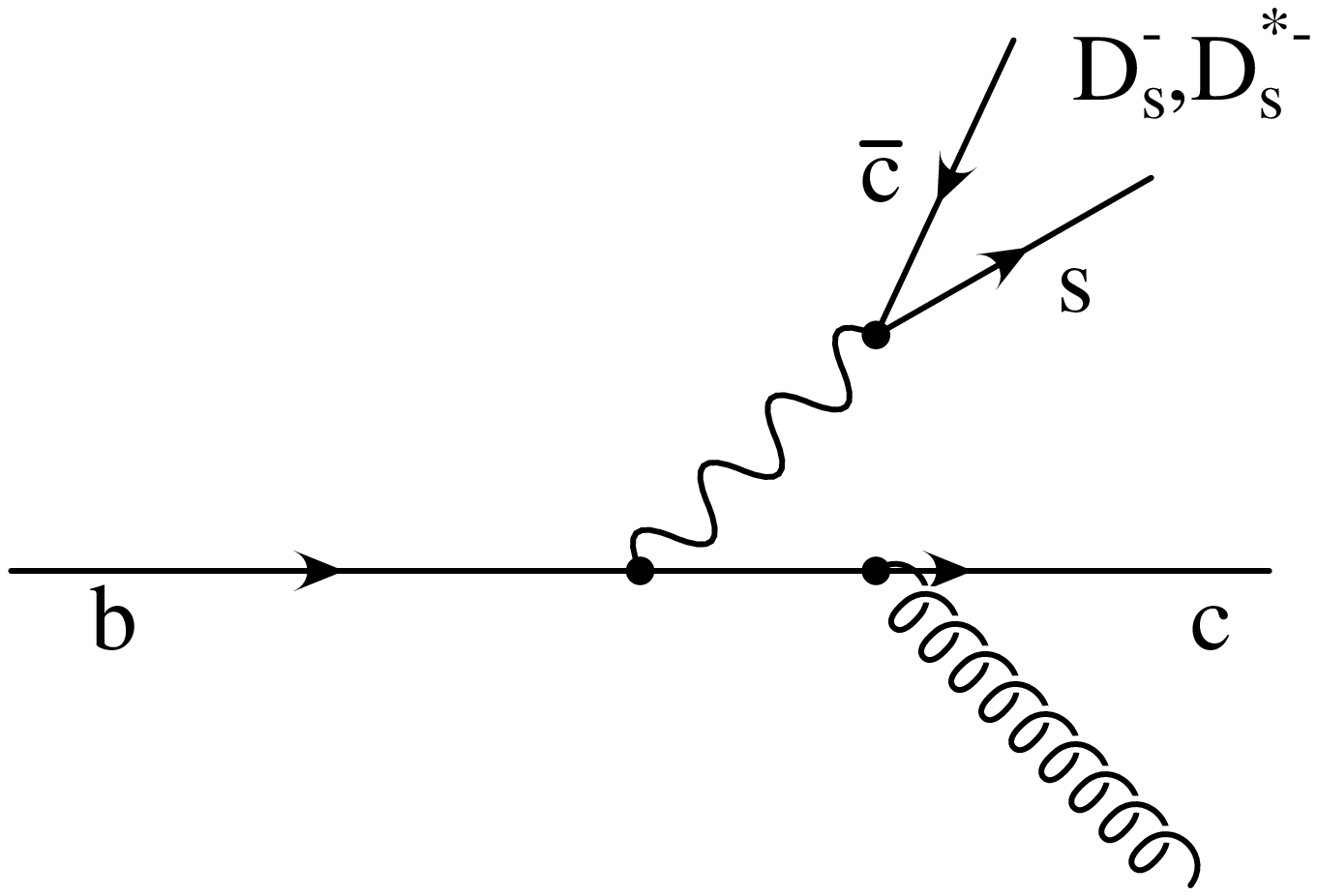,scale=0.4}
\caption{\label{fig2}Leading order Born term contribution (a) and $O(\alpha_s)$
contributions (b, c, d) to $b\rightarrow c+(D_s^-,D_s^{*-})$.}
\end{figure}

As explained in Sec.~2, the decay $\Lam_b\rightarrow X_c+(D_s^-,D_s^{*-})$
involves seven structure functions which can be resolved by an angular
analysis of the decay products. We begin by writing down the leading order
Born term contributions given by the quark level transition
$b\rightarrow c+(D_s^-,D_s^{*-})$ (see Fig.~\ref{fig2}a). For the partial
helicity rates one obtains
\begin{equation}
\label{gammaborn} 
  \Gamma_i^{\rm Born} (b^{(\uparrow)} \rightarrow c+D_s^{(*)-}) = 
  \frac{G_F^2}{8 \pi} |V_{bc} V_{cs}^{*}|^2
  f_{D_s^{(*)}}^2 m_b^2\,p_{D_s^{(*)}}\,a_1^2\,B_i,
\end{equation}
where
\begin{eqnarray}
\label{bornrates} 
  B_{S^{\phantom P}} = B_{L^{\phantom P}} & = &
  (1-y^2)^2-x^2 (1+y^2), \label{born1} \nonumber\\[3mm]
  B_{U^{\phantom P}} & = &
  2 x^2 (1-x^2+y^2), \label{born3} \nonumber\\[3mm]
  B_{U+L} & = & (1-y^2)^2+x^2(1+y^2-2x^2), \label{born4} \nonumber\\[3mm]
  B_{S^P} = B_{L^P} & = & \sqrt{\lam} (1-y^2), \label{born5} \\[3mm]
  B_{U^P} & = &-2 x^2 \sqrt{\lam}, \label{born6} \nonumber\\[3mm]
  B_{(U+L)^P} & = & \sqrt{\lam} (1-y^2-2x^2),\label{born7} \nonumber\\
  B_{I^P} & = &-\frac{1}{\sqrt{2}} x \sqrt{\lam}, \label{born8} \nonumber
\end{eqnarray}
and where $x=m_{D_s^{(*)}}/m_b$ and $y=m_c/m_b$. The kinematical factor $\lam$
is defined by $\lam=1+x^4+y^4-2(x^2+y^2+x^2y^2)$ such that
$p_{D_s^{(*)}}=\frac12m_b\,\lam^{1/2}$. In Eq.~(\ref{gammaborn}), $f_{D_s}$
and $f_{D^*_s}$ denote the pseudoscalar and vector meson coupling constants
defined by $\langle D_s^-|A^\mu|0\rangle=if_{D_s}p^\mu_{D_s}$ and
$\langle D_s^{*-}|V^\mu|0\rangle=f_{D_s^{*}}m_{D_s^{*}}\epsilon^{*\mu}$,
respectively. The Kobayashi--Maskawa matrix element is denoted by $V_{q_1q_2}$,
and the $p_{D_s}$ and $p_{D_s^*}$ are the magnitude of the three-momenta of
the $D_s$ and $D_s^*$ in the $b$ rest system. The parameter $a_1$ is related
to the Wilson coefficients of the renormalized current-current interaction and
is obtained from a combined fit of several decay modes ($|a_1|=1.00\pm 0.06$)
\cite{aleksan00}. Note that the structural similarity of the unpolarized and
polarized rate formulae for the decay into $D_s$ and into the longitudinal
$D_s^*$ is an accident of the Born term calculation and does not persist e.g.\
at higher orders of $\alpha_s$, or for the nonperturbative contributions to
the unpolarized longitudinal rate into $D_s^*$ to be written down later on.

$\Gamma_S^{\rm Born}$ and $\Gamma_{U+L}^{\rm Born}$ determine the total  
$\Lam_b\rightarrow X_c+D_s^-$ and $\Lam_b\rightarrow X_c+D_s^{*-}$ rates at
the Born term level, respectively. Using $f_{D_s}=230\mbox{ MeV}$ and
$f_{D_s^*}=280\mbox{ MeV}$ as in \cite{aleksan00},
$\tau_{\Lam_b}=1.23\mbox{ ps}$, $V_{bc}=0.04$, $V_{cs}=0.974$ and the central
value for $a_1$, one arrives at
\begin{equation} 
 \mbox{BR}_{\Lambda_b \rightarrow X_c+ D_s^{- } } \cong 2.5 \%, \qquad
 \mbox{BR}_{\Lambda_b \rightarrow X_c+D_s^{*-}} \cong 5.2 \%.
\end{equation}

While the semi-inclusive $\Lam_b\rightarrow X_c+(D_s^-,D_s^{*-})$ rates have
not been measured yet, a comparison of the Born term prediction with data on
the corresponding mesonic decay $\bar{B^0}\rightarrow X_c+(D_s^-,D_s^{*-})$ is
meaningful because the Born level predictions for both processes are identical.
Allowing for the factor $\tau_{\Lam_b}/\tau_B\approx 0.77$ and summing up the
$D_s$ and $D_s^*$ modes, one arrives at a branching ratio of $10\%$ which is
consistent with the measured value
$\mbox{BR}(B\rightarrow X+D_s^\pm)=(10.0\pm 2.5)\%$ \cite{pdg02} if one
assumes that the above two rates saturate the semi-inclusive rate into
$D_s^\pm$.

\section{\pf $ O(\alpha_s) $ radiative corrections}

Next we turn to the $O(\alpha_s)$ radiative corrections. As is evident from
Fig.~\ref{fig2}, the radiative gluon corrections connect only to the $b$ and
$c$ legs of the parton decay process $b\rightarrow c+(D_s,D_s^*)$ because of
the conservation of colour (see Fig.~2b, 2c and 2d). The radiative corrections
for the seven structure functions are thus identical to the corresponding
radiative corrections calculated in \cite{fgkm02} where the process
$t\rightarrow W^++b$ was considered (including the scalar case) keeping
$m_b\neq 0$ \footnote{In the unpolarized case the total $O(\alpha_s)$
correction to the spin $1$ piece of the weak current keeping both quark masses
finite had been calculated before in \cite{hokim84,jk89,czarnecki90,ds91}.
The $O(\alpha_s)$ corrections to the (unpolarized) spin $0$ piece of the weak
current can also be deduced from the calculations of \cite{hokim84,cjk95}.}. 

For the $O(\alpha_s)$ radiative corrections one has to calculate the square 
of the tree-graph amplitudes Figs.~2c and 2d, and the one-loop contribution
Fig.~2b. We concentrate on the tree--graph contribution given by the squares 
of the tree-graph amplitudes Fig.~2b and 2c which will be denoted by 
${\cal H}^{\mu \nu}(tree)$. The $b \rightarrow c$ hadron 
tensor can be obtained from the corresponding $t \rightarrow b$ hadron 
tensor given in
\cite{fgkm02} by the replacements $p_t \rightarrow p_b$ and 
$p_b \rightarrow p_c$ \footnote{We take the opportunity to correct two sign
typos in the corresponding $t \rightarrow b$ expression in \cite{fgkm02}}.
One obtains

\begin{eqnarray} 
 \label{Hadrontensor}
\lefteqn{{\cal H}^{\mu \nu}(tree) \ = \-4 \pi \alpha_s\,C_F \,
  \frac{8}{(k \!\cdot\! p_b)(k \!\cdot\! p_c)} \Bigg\{ \! -
  \frac{k \!\cdot\! p_b}{k \!\cdot\! p_c} \Big[ m_c^2 \Big(
  k^{\mu}\,\bar{p}_b^{\nu}+k^{\nu}\,\bar{p}_b^{\mu} -
  k \!\cdot\! \bar{p}_b\,g^{\mu \nu} \Big)+}
  \\ & & \hspace{-0.4cm} \rule{0mm}{6mm}+i \Big(
  \epsilon^{\alpha \beta \mu \nu}\,(p_c \!-\! k) \!\cdot\! \bar{p}_b -
  \epsilon^{\alpha \beta \gamma \nu}(p_c \!-\! k)^{\mu}\,\bar{p}_{b,\gamma}+
  \epsilon^{\alpha \beta \gamma \mu} (p_c \!-\! k)^{\nu}\,\bar{p}_{b,\gamma} 
  \Big) k_{\alpha}\,p_{c,\beta} \Big] +
  \nonumber \\ & & \hspace{-0.4cm} \rule{0mm}{6mm} +
  \frac{k \!\cdot\! p_c}{k \!\cdot\! p_b} \Big[ (\bar{p_b} \cdot p_b)
  \Big( k^{\mu}\,p_c^{\nu}+k^{\nu}\,p_c^{\mu} -
  k \!\cdot\! p_c\,g^{\mu \nu} -
  i\,\epsilon^{\alpha \beta \mu \nu} k_{\alpha}\,p_{c,\beta} \Big) +
  \nonumber \\ & & \hspace{-0.4cm} \rule{0mm}{6mm}-
  (\bar{p}_b \!\cdot\! k) \Big( (p_b \!-\! k)^{\mu}\,p_b^{\nu} +
  (p_b \!-\! k)^{\nu}\,p_c^{\mu}-(p_b \!-\! k) \!\cdot\! p_c\,g^{\mu \nu} -
  i\,\epsilon^{\alpha \beta \mu \nu}
  (p_b \!-\! k)_{\alpha}\,p_{c,\beta} \Big) \Big] +
  \nonumber \\ & & \hspace{-0.4cm} \rule{0mm}{6mm}  -
  (\bar{p}_b \!\cdot\! p_c) \Big( k^{\mu}\,p_c^{\nu} +
  k^{\nu}\,p_c^{\mu}-k \!\cdot\! p_c\,g^{\mu \nu} -
  i\,\epsilon^{\alpha \beta \mu \nu} k_{\alpha}\,p_{c,\beta} \Big) +
  (p_b \!\cdot\! p_c) \Big( k^{\mu}\,\bar{p}_b^{\nu} +
  k^{\nu}\,\bar{p}_b^{\mu}-k \!\cdot\! \bar{p}_b\,g^{\mu \nu} \Big)
  \!+\kern-5pt
  \nonumber \\ & & \hspace{-0.4cm} \rule{0mm}{6mm} -
  (k \!\cdot\! p_c) \Big( p_b^{\mu}\,\bar{p}_b^{\nu} +
  p_b^{\nu}\,\bar{p}_b^{\mu}-p_b \!\cdot\! \bar{p}_b\,g^{\mu \nu} \Big) +
  (k \!\cdot\! p_b) \Big( (p_c \!+\! k)^{\mu}\,\bar{p}_b^{\nu} \!+\!
  (p_c \!+\! k)^{\nu}\,\bar{p}_b^{\mu} \!-\! (p_c \!+\! k) \!\cdot\! \bar{p}_b
 \,g^{\mu \nu} \Big)+\kern-5pt
  \nonumber \\ & & \hspace{-0.4cm} \rule{0mm}{6mm} +
  (k \!\cdot\! \bar{p}_b)
  \Big( 2 p_c^{\mu}\,p_c^{\nu}-p_c \!\cdot\! p_c g^{\mu \nu} \Big) +
  i \Big(\epsilon^{\alpha \beta \mu \nu}\,(k \!\cdot\! \bar{p}_b) +
  \epsilon^{\alpha \beta \gamma \mu}\,k^{\nu} \bar{p}_{b,\gamma} -
  \epsilon^{\alpha \beta \gamma \nu}\,k^{\mu} \bar{p}_{b,\gamma} \Big)
  p_{c,\alpha}\,p_{b,\beta} +
  \nonumber \\ & & \hspace{-0.4cm} \rule{0mm}{6mm} +
  i \Big(\epsilon^{\alpha \beta \mu \nu}\,(p_b \!\cdot\! \bar{p}_b) +
  \epsilon^{\alpha \beta \gamma \mu}\,p^{\nu}_b \bar{p}_{b,\gamma} -
  \epsilon^{\alpha \beta \gamma \nu}\,p^{\mu}_b \bar{p}_{b,\gamma} \Big)
  k_{\alpha}\,p_{c,\beta} \Bigg\}+B^{\mu \nu} \cdot \Delta_{SGF}
  \nonumber
 \end{eqnarray}
 \begin{equation} 
  \Delta_{SGF} :=-4 \pi \alpha_s\,C_F
  \Big( \frac{m_c^2}{(k \!\cdot\! p_c)^2}+\frac{m_b^2}{(k \!\cdot\! p_b)^2} -
  2 \frac{p_c \!\cdot\! p_b }{(k \!\cdot\! p_c)(k \!\cdot\! p_b)} \Big)
 \end{equation}

 \noindent where $ k $ is the 4-momentum of the emitted gluon. The
polarization of the bottom quark is taken into account by introducing
the short-hand notation $ \bar{p}_b = p_b-m_b s_b $. We have found
it convenient to split the tree-graph hadron tensor into an infrared
(IR) finite piece and an IR divergent piece given by the usual soft-gluon
factor $\Delta_{SGF}$ multiplied by the Born term tensor $B^{\mu \nu}$

\begin{equation} 
 \label{bornspur}
  B^{\mu \nu} = 2 ( \bar{p}_b^{\nu} p_c^{\mu}+\bar{p}_b^{\mu} p_c^{\nu} -
  g^{\mu \nu}\,\bar{p}_b \!\cdot\! p_c+i \epsilon^{\mu \nu \alpha \beta}
  p_{c,\alpha} \bar{p}_{b,\beta}).
 \end{equation}   
In this way the IR singularity is isolated in the universal function
$\Delta_{SGF}$ which can be integrated by introducing a gluon mass 
regulator to regularize the IR singularity. The ensuing logarithmic gluon 
mass singularity is cancelled by the corresponding gluon mass singularity
occurring in the loop contribution (see e.g.\cite{fgkm02}). 

The phase-space
integration of the IR convergent piece can be done without a gluon
mass regulator. One first projects the convergent piece of the tree-graph 
tensor (\ref{Hadrontensor}) onto the seven helicity structure functions 
Eq.(8) and then does the phase-space integration in the sequential
order (i) $k_0$ (gluon energy) (ii) $q_0$ (energy of the off-shell $W^-$).

The final $O(\alpha_s)$ answer including the one-loop contribution can be 
written in a very compact way by introducing combinations of
dilogarithmic functions ${\cal A}$ and ${\cal N}_{0,..,4}$.
The contribution denoted by ${\cal A}$ is part of the finite remainder
of the Born term
type one-loop contribution plus the soft gluon contribution. The
combinations ${\cal N}_{0,..,4}$ appear when integrating those 
helicity structure functions $H_{S^P}, H_U, H_{U^P}, H_{L^P}, H_{L^P}$ and
$H_{I^P}$ that are not associated with the total rate.
All these 
functions are defined after Eq.(\ref{defgammaip}). We mention that we have 
now been able to present our results on the radiatively corrected
structure functions in a much more compact form than thought possible
when we wrote up \cite{fgkm02}.  

We shall
present our $O(\alpha_s)$ results in a form where the respective Born terms
$\Gamma_i^{(0)}$ are factored out from the $O(\alpha_s)$ result. Including the
Born term and the nonperturbative $O(1/m_b^2)$ contributions to be discussed
in Sec.~4 we write $\hat\Gamma_i:=\Gamma_i/\Gamma_S^{\rm Born}$ and
$\hat\Gamma_i^{\rm Born}:=\Gamma_i^{\rm Born}/\Gamma_S^{\rm Born}$ for
$i=S, S^P$, and $\hat\Gamma_i:=\Gamma_i/\Gamma_{U+L}^{\rm Born}$ and
$\hat\Gamma_i^{\rm Born}:=\Gamma_i^{\rm Born}/\Gamma_{U+L}^{\rm Born}$ for
$i=U, L, U+L, U^P, L^P, U^P+L^P, I^P$. One has
\begin{equation} 
\label{altogether}
  \hat\Gamma_i = \hat\Gamma_i^{\rm Born} 
  \bigg( 1+C_F \frac{\alpha_s}{4 \pi} \tilde{\Gamma}_i +
  a_i^K\,K_b+a_i^\epsilon\,\epsilon_b \bigg).
\end{equation}
$K_b$ is the expectation value of the kinetic energy of the heavy quark in the
$\Lam_b$ baryon and $\epsilon_b$ parametrizes the spin-dependent contribution
of the heavy quark in the $\Lam_b$ baryon \cite{Manohar:1993qn}.

To begin with, we list the reduced $O(\alpha_s)$ rates $\tilde\Gamma_i$. For
the reduced unpolarized and polarized scalar spin $0$ rates $\tilde\Gamma_S$
and $\tilde\Gamma_{S^P}$ we obtain
\begin{eqnarray} 
\label{defgammas}
\tilde{\Gamma}_S &=& {\cal{A}}+\frac{1}{\sqrt{\lambda}}\,\Big[ \lambda
     +x^2 \left( 1-x^2+y^2 \right) \Big]^{- 1} \nonumber \\ [1.5mm]
     && \times \; \Bigg\{ \frac{2}{x^2}\,\Big[
     \left( 1-x^2 \right) \left( 2-x^2 \right)
     -\left( 6+4\,x^2+5\,x^4 \right) y^2 \nonumber \\ [1.5mm]
     && \hspace{1cm}+\; \left( 6+7\,x^2
     \right) y^4-2\,y^6 \Big]\,\sqrt{\lambda}\,\ln (y)
     \nonumber \\ [1.5mm]
     &&+\; 8\,\Big[ 1-x^2-\left( 2+x^2 \right) y^2+y^4 \Big] \,
     \sqrt{\lambda}\,\ln \left( \frac{x\,y}{\lambda} \right)
     \nonumber \\ [1.5mm]
     &&-\; \frac{1}{x^2}\,\Big[ \left( 1-x^2 \right)^2
     \left( 2+3\,x^2 \right)-\left( 8-3\,x^2+4\,x^4-3\,x^6
     \right) y^2 \nonumber \\ [1.5mm]
     && \hspace{1cm}+\; 3 \left( 4+5\,x^2 \right) y^4
     -\left( 8+5\,x^2 \right) y^6+2\,y^8 \Big]\,\ln (w_1)
     \nonumber \\ [1.5mm]
     &&+\; 8 \left( 1-y^2 \right) \Big[ 1-x^2-\left( 2+x^2
     \right) y^2+y^4 \Big]\,\ln (\eta) \nonumber \\ [1.5mm]
     &&+\; 3\,\sqrt{\lambda}\,\Big[ 3 \left( 1-x^2 \right)
     -\left( 10+3\,x^2 \right) y^2+3\,y^4 \Big] \Bigg\}
\end{eqnarray}
\begin{eqnarray} 
\label{defgammasp}
\tilde{\Gamma}_{S^P} &=& {\cal{A}}+\frac{1}{\lambda} \,
     \Bigg\{ 4\,\Big[ 2+x^4-\left( 3 +2\,x^2
     \right) y^2+y^4 \Big]\,{\cal{N}}_0+4\,\sqrt{\lambda}
     \left( 1-x^2+y^2 \right) {\cal{N}}_4 \nonumber \\ [1.5mm]
     &&+\; \frac{2}{x^2}\,\frac{1}{1-y^2}\,\Big[ 2-x^2
     -\left( 4+5\,x^2 \right) y^2+2\,y^4 \Big]\,\lambda\,\ln (y)
     +8\,\lambda\,\ln \left( \frac{x\,y}{\lambda} \right)
     \nonumber \\ [1.5mm]
     &&-\; \frac{\sqrt{\lambda}}{x^2}\,\Big[ 2-9\,x^2+x^4
     -\left( 4+3\,x^2 \right) y^2+2\,y^4 \Big]\,\ln (w_1)
     +8\,\lambda\,\ln \left[ \frac{(1+x)^2-y^2}{x}
     \right] \nonumber \\ [1.5mm]
     &&+\; 4\,\Big[ \left( 1-x^2 \right) \left( 5-2\,x^2 \right)
     +2 \left( 2-x^2 \right) y^2 \Big]\,\ln \left( \frac{1-x}{y}
     \right) \nonumber \\ [1.5mm]
     &&-\; \Big[ (1-x)^2-y^2 \Big] \left( 11-6\,x-7\,x^2
     +7\,y^2 \right) \Bigg\}
\end{eqnarray}
The other variables and functions appearing in Eqs.~(\ref{defgammas}) and
(\ref{defgammasp}) are 
explained at the end of this section. For the five unpolarized and polarized
reduced spin $1$ rates $\tilde\Gamma_i$ ($i= U, L, U^P, L^P, I^P$) we obtain
\begin{eqnarray} 
\label{defgammau}
\tilde{\Gamma}_U &=& {\cal{A}}+\frac{1}{\sqrt{\lambda}} \,
     \frac{1}{1-x^2+y^2}\,\Bigg\{-\; 4 \left( 7+x^2-y^2 \right)
     {\cal{N}}_1 \nonumber \\ [1.5mm]
     &&-\; \frac{2}{x}\,\Big[ (1-x)^2-y^2 \Big]\,\Big[
     (1-x)\,(5+x)+y^2 \Big]\,{\cal{N}}_2 \nonumber \\ [1.5mm]
     &&-\; \frac{2}{x}\,\Big[ (1+x)^2-y^2 \Big]\,\Big[
     (1+x)\,(5-x)+y^2 \Big]\,{\cal{N}}_3 \nonumber \\ [1.5mm]
     &&-\; \frac{2}{x^2} \left( 1-x^2+y^2 \right)
     \left( 1-2\,x^2-y^2 \right) \sqrt{\lambda}\,\ln (y)
     +8 \left( 1-x^2+y^2 \right) \sqrt{\lambda}\,\ln \left(
     \frac{x\,y}{\lambda} \right)
     \nonumber \\ [1.5mm]
     &&+\; \frac{1}{x^2}\,\Big[ \left( 1-x^2 \right)^2
     \left( 1-6\,x^2 \right) \nonumber \\ [1.5mm]
     && \hspace{1cm}-\; \left( 1+4\,x^2-3\,x^4 \right) y^2-\left(
     1+2\,x^2 \right) y^4+y^6 \Big]\,\ln (w_1)
     \nonumber \\ [1.5mm]
     &&+\; 4\,\Big[ 7+3\,x^2
     -\left( 4-5\,x^2 \right) y^2-3\,y^4
     \Big]\,\ln (\eta)-\sqrt{\lambda}
     \left( 19+x^2-5\,y^2 \right) \Bigg\}
\end{eqnarray}
\begin{eqnarray} 
\label{defgammal}
\tilde{\Gamma}_L &=& {\cal{A}}+\frac{1}{\sqrt{\lambda}}\,\Big[
     \lambda+x^2 \left( 1-x^2+y^2 \right) \Big]^{- 1} \,
     \Bigg\{ 8\,x^2 \left( 7+x^2-y^2 \right) {\cal{N}}_1
     \nonumber \\ [1.5mm]
     &&+\; 4\,x\,\Big[ (1-x)^2-y^2 \Big]\,\Big[ (1-x)\,(5+x)
     +y^2 \Big]\,{\cal{N}}_2 \nonumber \\ [1.5mm]
     &&+\; 4\,x \; \Big[ (1+x)^2-y^2 \Big]\,\Big[ (1+x)\,(5-x)
     +y^2 \Big]\,{\cal{N}}_3 \nonumber \\ [1.5mm]
     &&+\; 2\,\Big[ 1-x^2-\left( 4+3\,x^2 \right) y^2+3\,y^4
     \Big]\,\sqrt{\lambda}\,\ln (y) \nonumber \\ [1.5mm]
     &&+\; 8\,\Big[ 1-x^2-\left( 2+x^2 \right) y^2+y^4 \Big] \,
     \sqrt{\lambda}\,\ln \left( \frac{x\,y}{\lambda} \right)
     \nonumber \\ [1.5mm]
     &&-\; \Big[ 5 \left( 1-x^2 \right)^2-\left( 3+20\,x^2-x^4
     \right) y^2+\left( 9-2\,x^2 \right) y^4+y^6 \Big]\,\ln (w_1)
     \nonumber \\ [1.5mm]
     &&+\; 8 \left( 1+x^2-y^2 \right) \Big[ 1-7\,x^2-\left(
     2+x^2 \right) y^2+y^4 \Big]\,\ln (\eta) \nonumber \\ [1.5mm]
     &&+\; \sqrt{\lambda}\,\Big[ 5+47\,x^2-4\,x^4-\left(
     22+x^2 \right) y^2+5\,y^4 \Big] \Bigg\}
\end{eqnarray}
\begin{eqnarray} 
\label{defgammaul}
\tilde{\Gamma}_{U+L} &=& {\cal{A}}+\frac{1}{\sqrt{\lambda}} \,
     \Big[ \lambda+3\,x^2 \left( 1-x^2+y^2 \right) \Big]^{- 1}
     \nonumber \\ [1.5mm]
     && \times \; \Bigg\{ -\,2\,\Big[ \left( 1-x^2 \right)
     \left( 1-4\,x^2 \right)+\left( 4+x^2 \right) y^2-5\,y^4
     \Big]\,\sqrt{\lambda}\,\ln (y) \nonumber \\ [1.5mm]
     &&+\; 8\,\Big[ \left( 1-x^2 \right) \left( 1+2\,x^2 \right)
     -\left( 2-x^2 \right) y^2+y^4 \Big]\,\sqrt{\lambda}\,\ln
     \left( \frac{x\,y}{\lambda} \right) \nonumber \\ [1.5mm]
     &&-\; \Big[ 3 \left( 1-x^2 \right)^2 \left( 1+4\,x^2 \right)
     \nonumber \\ [1.5mm]
     && \hspace{1cm}-\; \left( 1+12\,x^2+5\,x^4 \right) y^2
     +\left( 11+2\,x^2 \right) y^4-y^6 \Big]\,\ln (w_1)
     \nonumber \\ [1.5mm]
     &&+\; 8 \left( 1-y^2 \right) \Big[ 1+x^2-4\,x^4
     -\left( 2-x^2 \right) y^2+y^4 \Big]\,\ln (\eta)
     \nonumber \\ [1.5mm] 
     &&+\; \sqrt{\lambda}\,\Big[ 5+9\,x^2-6\,x^4
     -\left( 22-9\,x^2 \right) y^2+5 y^4 \Big] \Bigg\}
\end{eqnarray}
\begin{eqnarray} 
\label{defgammaup}
\tilde{\Gamma}_{U^P} &=& {\cal{A}}+\frac{1}{\lambda} \,
     \Bigg\{ 4\,\Big[ 11+3\,x^2+x^4-2 \left(
     3+x^2 \right) y^2+y^4 \Big]\,{\cal{N}}_0+4\,\sqrt{\lambda}
     \left( 1-x^2+y^2 \right) {\cal{N}}_4 \nonumber \\ [1.5mm]
     &&-\; \frac{2}{x^2} \left( 1-2\,x^2-y^2 \right)
     \lambda\,\ln (y)+8\,\lambda\,\ln \left( \frac{x\,y}{\lambda}
     \right) \nonumber \\ [1.5mm]
     &&+\; \frac{\sqrt{\lambda}}{x^2}\,\Big[ 7+21\,x^2+2\,x^4
     -\left( 8+3\,x^2 \right) y^2+y^4 \Big] \ln (w_1)
     \nonumber \\ [1.5mm]
     &&+\; 8\,\lambda\,\ln \left[ \frac{(1+x)^2-y^2}{x} \right]
     \nonumber \\ [1.5mm]
     &&+\; \frac{4}{x^2}\,\Big[ \left( 1-x^2 \right)
     \left( 3+14\,x^2-2\,x^4 \right) \nonumber \\ [1.5mm]
     && \hspace{1cm}-\; \left( 6-7\,x^2-x^4
     \right) y^2+\left( 3-x^2 \right) y^4 \Big]\,\ln \left(
     \frac{1-x}{y} \right) \nonumber \\ [1.5mm]
     &&+\; \frac{1}{x}\,\Big[ (1-x)^2-y^2 \Big]\,\Big[
     12-55\,x+6\,x^2-x^3-3 \left( 4+x \right) y^2 \Big]
     \Bigg\}
\end{eqnarray}
\begin{eqnarray} 
\label{defgammalp}
\tilde{\Gamma}_{L^P} &=& {\cal{A}}+\frac{1}{\lambda}\,\frac{1}{1-y^2}
     \nonumber \\ [1.5mm]
     && \times \; \Bigg\{ 4\,\Big[ 2+22\,x^2+11\,x^4
     -\left( 5+12\,x^2+x^4 \right) y^2+2 \left( 2+x^2 \right) y^4
     -y^6 \Big]\,{\cal{N}}_0 \nonumber \\ [1.5mm]
     &&+\; 4 \left( 1-y^2 \right) \sqrt{\lambda} \left( 1-x^2+y^2
     \right) {\cal{N}}_4 \nonumber \\ [1.5mm]
     &&+\; 2 \left( 1-3\,y^2 \right)
     \lambda\,\ln (y)+8 \left( 1-y^2 \right) \lambda\,\ln
     \left( \frac{x\,y}{\lambda} \right) \nonumber \\ [1.5mm]
     &&+\; \sqrt{\lambda}\,\Big[ 17+53\,x^2-\left( 18+x^2
     \right) y^2+y^4 \Big]\,\ln (w_1) \nonumber \\ [1.5mm]
     &&+\; 8\,\lambda \left( 1-y^2 \right) \ln \left[
     \frac{(1+x)^2-y^2}{x} \right] \nonumber \\ [1.5mm]
     &&+\; 4\,\Big[ \left( 1-x^2 \right) \left( 11+24\,x^2 \right)
     -\left( 13-15\,x^2 \right) y^2+2\,y^4 \Big]\,\ln \left(
     \frac{1-x}{y} \right) \nonumber \\ [1.5mm]
     &&-\; \Big[ (1-x)^2-y^2 \Big] \\ [1.5mm]
     && \hspace{1cm}\times \; \Big[ 15-22\,x+105\,x^2
     -24\,x^3+4\,x^4-\left( 12-22\,x+x^2 \right) y^2
     -3\,y^4 \Big] \Bigg\} \nonumber
\end{eqnarray}
\begin{eqnarray} 
\label{defgammauplp}
\tilde{\Gamma}_{U^P+L^P} &=& {\cal{A}}+\frac{1}{\lambda} \,
     \frac{1}{1-2\,x^2-y^2}\,\Bigg\{ 4\,\Big[
     2+5\,x^4-2\,x^6-\left( 5-3\,x^4 \right) y^2
     +4\,y^4-y^6 \Big]\,{\cal{N}}_0 \nonumber \\ [1.5mm]
     &&+\; 4\,\sqrt{\lambda} \left( 1-x^2+y^2 \right) \left(
     1-2\,x^2-y^2 \right) {\cal{N}}_4 \nonumber \\ [1.5mm]
     &&+\; 2\,\lambda \left( 3-4\,x^2-5\,y^2 \right) \ln (y)
     +8\,\lambda \left( 1-2\,x^2-y^2 \right)\,\ln \left(
     \frac{x\,y}{\lambda} \right) \nonumber \\ [1.5mm]
     &&+\; \left( 3-x^2+y^2 \right) \left( 1+4\,x^2-y^2 \right)
     \sqrt{\lambda}\,\ln (w_1) \nonumber \\ [1.5mm]
     &&+\; 8\,\lambda \left( 1-2\,x^2-y^2 \right) \ln \left[
     \frac{(1+x)^2-y^2}{x} \right] \nonumber \\ [1.5mm]
     &&+\; 4\,\Big[ \left( 1-x^2 \right) \left( 5-4\,x^2+4\,x^4
     \right) \nonumber \\ [1.5mm]
     && \hspace{1cm}-\; \left( 1-x^2+2\,x^4 \right) y^2
     -2 \left( 2-x^2 \right) y^4 \Big]\,\ln \left( \frac{1-x}{y}
     \right) \nonumber \\ [1.5mm]
     &&-\; \left[ (1-x)^2-y^2 \right] \\ [1.5mm]
     && \hspace{1cm} \times \; \Big[ 15+2\,x-5\,x^2-12\,x^3
     +2\,x^4-\left( 12+2\,x+7\,x^2 \right) y^2-3\,y^4 \Big]
     \Bigg\} \nonumber
\end{eqnarray}
\begin{eqnarray} 
\label{defgammaip}
\tilde{\Gamma}_{I^P} &=& {\cal{A}}+\frac{1}{\lambda} \,
     \Bigg\{ 2\,\Big[ 7+15\,x^2+4\,x^4
     -\left( 11+8\,x^2 \right) y^2+4\,y^4 \Big]\,{\cal{N}}_0
     \nonumber \\ [1.5mm]
     &&+\; 4\,\sqrt{\lambda} \left( 1-x^2+y^2 \right) {\cal{N}}_4
     \nonumber \\ [1.5mm]
     &&-\; \frac{1}{x^2} \left( 1-3\,x^2-y^2 \right) \lambda \ln (y)
     +8\,\lambda\,\ln \left( \frac{x\,y}{\lambda} \right)
     \nonumber \\ [1.5mm]
     &&+\; \frac{\sqrt{\lambda}}{2\,x^2}\,\Big[ 1+30\,x^2+21\,x^4
     -2 \left( 1+11\,x^2 \right) y^2+y^4 \Big]\,\ln (w_1)
     \nonumber \\ [1.5mm]
     &&+\; 8\,\lambda\,\ln \left[
     \frac{(1+x)^2-y^2}{x} \right] \nonumber \\ [1.5mm]
     &&+\; 2\,\Big[ \left( 1-x^2 \right) \left( 21+5\,x^2 \right)
     -\left( 11-15\,x^2 \right) y^2-4\,y^4 \Big]\,\ln \left(
     \frac{1-x}{y} \right) \nonumber \\ [1.5mm]
     &&-\; 2\,\Big[ (1-x)^2-y^2 \Big]
     \left( 12-7\,x+12\,x^2-9\,y^2 \right) \Bigg\}
\end{eqnarray}


As mentioned before the contribution denoted by ${\cal A}$ is that part 
of the finite remainder of 
the Born term type one-loop contribution plus the soft gluon contribution 
which contains dilogs and products or squares of logs . It is given by 
\begin{eqnarray} 
{\cal{A}} &=& \frac{2}{\sqrt{\lambda}} \left( 1-x^2+y^2 \right) \Bigg\{
      -\,4\,\mbox{Li}_2 (1-w_1) +\,4\,\mbox{Li}_2 (1-w_2)
      -\,4\,\mbox{Li}_2 (1-w_3) \\ [3mm]
     &&-\; \ln (w_1)\,\ln \left( \frac{\lambda^2\,w_3}{x^2\,y^3}
     \right)-\frac{1}{2}\,\ln^2 (w_1)
     +\ln \left[ \frac{1}{2} \left( 1-x^2+y^2+\sqrt{\lambda} \right)
     \right] \ln \left( w_2\,w_3 \right) \Bigg\}\,. \nonumber 
\end{eqnarray}
The functions ${\cal N}_{0,..,4}$ are defined by
\begin{eqnarray}
{\cal{N}}_0 &=& \mbox{Li}_2 \left( \frac{x}{\eta} \right)
    +\mbox{Li}_2 (x\,\eta)-2\,\mbox{Li}_2 (x) \nonumber \\ [6mm]
{\cal{N}}_1 &=& \mbox{Li}_2 (x\,\eta)
     -\mbox{Li}_2 \left( \frac{x}{\eta} \right)+2\,\ln (1-\eta\,x) \,
     \ln \left[ \frac{(\eta+1)\,x}{1+x} \right]+\ln \left(
     \frac{\eta}{\eta-x} \right) \ln \left[
     \frac{x^2\,(\eta-1)^2}{\eta\,(\eta-x)} \right] \nonumber \\ [6mm]
{\cal{N}}_2 &=& \mbox{Li}_2 \left[
     \frac{(\eta-1)\,x}{\eta-x} \right]+\mbox{Li}_2 \left[
     \frac{(\eta-1)\,x}{1-x} \right]-\frac{1}{2}\,\ln^2 (1-x)
     +\ln \left( \frac{\eta}{\eta-x} \right) \ln \left[
     \frac{(\eta-1)\,x}{\eta-x} \right] \nonumber \\ [3mm]
     &&+\; \ln (1-x)\,\ln \left( \frac{1-x}{\eta-x}
     \right)+\ln (1-\eta\,x)\,\ln \left[ \frac{(\eta+1)\,x}{1+x}
     \right] \nonumber \\ [6mm]
{\cal{N}}_3 &=& \mbox{Li}_2 \left(
     \frac{1-\eta\,x}{1+x} \right)-\mbox{Li}_2 \left[
     \frac{\eta-x}{\eta\,(1+x)} \right]-\frac{1}{2}\,\ln \left(
     \frac{\eta}{\eta-x} \right) \ln \left[
     \frac{\eta\,(\eta-1)^2\,(1+x)^2}{(\eta+1)^2\,(\eta-x)}
     \right] \nonumber \\ [6mm]
{\cal{N}}_4 &=& 4\,\mbox{Li}_2 \left(
     \frac{\eta\,\sqrt{\lambda}}{\eta-x} \right)-2\,\mbox{Li}_2
     \left[ \frac{(\eta-1)\,x}{1-x} \right]-2\,\mbox{Li}_2 \left[
     \frac{(\eta-1)\,x}{\eta-x} \right]+\mbox{Li}_2 \left(
     \frac{x}{\eta} \right)-\mbox{Li}_2 (x\,\eta) \nonumber \\ [3mm]
     &&-\; \ln^2 (1-x)+\ln \left( \frac{\eta}{\eta-x} \right) \ln
     \left[ \frac{\eta\,(\eta+1)^2}{\eta-x} \right]+2 \,
     \ln (1-\eta\,x)\,\ln \left[ \frac{(1-x)\,(\eta+1)}{\eta-x}
     \right]\,, \nonumber
\end{eqnarray}
where we use the abbreviations
\begin{eqnarray}
w_1 &=& \frac{1-x^2+y^2-\sqrt{\lambda}}{1-x^2+y^2+\sqrt{\lambda}}
\; , \; \; \; \; \;
w_2 = \frac{1+x^2-y^2-\sqrt{\lambda}}{1+x^2-y^2+\sqrt{\lambda}}
\; , \; \; \; \; \;
w_3 = \frac{1-x^2-y^2-\sqrt{\lambda}}{1-x^2-y^2+\sqrt{\lambda}}
     \nonumber \\ [5mm]
\eta &=& \frac{1+x^2-y^2+\sqrt{\lambda}}{2\,x}
\end{eqnarray}


\section{Nonpertubative contributions}

When one uses the operator product expansion in HQET one can determine the
nonpertubative corrections to the leading partonic $b\rightarrow c$ rate. The
nonpertubative corrections set in at $O(1/m_b^2)$ and arise from the kinetic
energy and the spin dependent piece of the heavy quark in the heavy baryon
\cite{Manohar:1993qn}. The strength of the kinetic and the spin dependent
piece are parametrized by the expectation values of the relevant operators
in the $\Lam_b$ system and are denoted by $K_b$ and $\epsilon_b$, respectively.
We have completely recalculated the nonpertubative contributions to the seven
partial rates and have found some errors in the calculation of \cite{bkp98}
which will be corrected in an Erratum to \cite{bkp98}. One has
\begin{eqnarray} 
  S:^{\phantom P} \; a_{S^{\phantom P}}^K & = &-1, \qquad 
  a_{S^{\phantom P}}^\epsilon \ = \ 0 \nonumber\\
  U:^{\phantom P} \; a_{U^{\phantom P}}^K & = &
 -\left( 1-\frac{8}{3}\, \frac{1}{1-x^2+y^2}\right), \qquad
  a_{U^{\phantom P}}^\epsilon \ = \ 0 \nonumber\\
  L:^{\phantom P} \; a_{L^{\phantom P}}^K & = &
 -\left( 1+\frac{16}{3}\,\frac{x^2}{(1-y^2)^2-x^2(1+y^2)}\right),
  \qquad a_{L^{\phantom P}}^\epsilon \ = \ 0 , \qquad\qquad \nonumber\\  
  U+L:^{\phantom P} \; a_{U+L}^K & = &-1, \qquad
  a_{U+L^{\phantom P}}^\epsilon \ = \ 0 \nonumber\\
  S^P: \; a_{S^P}^K & = &
  -(1+\frac{8}{3} \frac{x^2}{\lam}) , \qquad 
  a_{S^P}^\epsilon \ = \ 1 \\
  U^P: \; a_{U^P}^K & = & -(1+\frac{8}{3} \frac{x^2}{\lam}) \qquad
  a_{U^P}^\epsilon \ = \ 1 \nonumber\\
  L^P: \; a_{L^P}^K & = &-(1+\frac{8}{3} \frac{x^2}{\lam}) \qquad  
  a_{L^P}^\epsilon \ = \ 1 \nonumber\\
  (U+L)^P: \; a_{(U+L)^P} & = & -(1+\frac{8}{3} \frac{x^2}{\lam}) \qquad 
  a_{(U+L)^P}^\epsilon \ = \ 1 \nonumber\\
  I^P: \; a_{I^P}^K & = & \frac{2}{3} (1-4\,\frac{x^2}{\lam}) \qquad  
  a_{I^P}^\epsilon \ = \ 1 \nonumber
\end{eqnarray}
The nonperturbative contributions for $(U+L)$ and $(U+L)^P$ can be compared to
the corresponding $q^2$-distributions in semi-leptonic $b$-decays written down
in \cite{Korner:1998nc}. We find agreement. 

For our numerical evaluation we use $K_b=0.013$ for the mean kinetic energy of
the heavy quark in the $\Lam_b$ as in \cite{bkp98}. An estimate of the
spin-dependent parameter has been given in \cite{fn93} with the result
$\epsilon_b=-\frac23K_b$, based on an assumption that the contribution of
terms arising from double insertions of the chromomagnetic operator can be
neglected. A zero recoil sum rule analysis gives the constraint
$\epsilon_b\le-\frac23K_b$ \cite{kp94} which puts the estimate of \cite{fn93}
at the upper bound of the constraint. We use the value of \cite{fn93} keeping
in mind that the numerical value of $\epsilon_b$ could be reduced in more
realistic calculations.


\section{Numerical results}

Using $m_b=4.85\mbox{ GeV}$, $m_c=1.45\mbox{ GeV}$,
$m_{D_s}=1968.5\mbox{ MeV}$, $m_{D_s^*}=2112.4\mbox{ MeV}$ and
$\alpha_s(m_b)=0.2$ we obtain for $b\rightarrow c$
\begin{eqnarray}
  \hat\Gamma_{S^{\phantom P} \phantom{+ L}} & = & \phantom{+0.0000 \,}
  (1-0.0964-0.0130+0), \nonumber\\[2mm]
  \hat\Gamma_{U^{\phantom P} \phantom{+ L}} & = & \pp
  0.3541\,(1-0.1079+0.0255+0), \nonumber\\[2mm]
  \hat\Gamma_{L^{\phantom P} \phantom{+ L}} & = & \pp
  0.6459\,(1-0.1103-0.0341+0), \nonumber\\[2mm]
  \hat\Gamma_{U+L^{\phantom P}} & = & \phantom{+0.0000 \,}
  (1-0.1095-0.0130+0), \nonumber\\[2mm]
  \hat\Gamma_{S^P \phantom{+ L}} & = & \pp
  0.9884\,(1-0.1027-0.0245-0.0087), \\[2mm]
  \hat\Gamma_{U^P \phantom{+ L}} & = & -
  0.2646\,(1-0.0616-0.0276-0.0087), \nonumber\\[2mm]
  \hat\Gamma_{L^P \phantom{+ L}} & = & \pp
  0.6351\,(1-0.1043-0.0276-0.0087), \nonumber\\[2mm]
  \hat\Gamma_{(U+L)^P{\phantom P}} & = & \pp
  0.3705\,(1-0.1348-0.0276-0.0087), \nonumber\\[2mm]
  \hat\Gamma_{I^P \phantom{+ L}} & = & -
  0.2148\,(1-0.0876+0.0059-0.0087). \nonumber
\end{eqnarray}
The four entries in the round brackets correspond to the Born term
contribution, the $O(\alpha_s)$ corrections, and the nonperturbative kinetic
and spin-dependent corrections in that order, as specified in
Eq.~(\ref{altogether}).

The reduction of the partial rates from the radiative corrections scatter
around $10\%$, where the reduction is largest for $\hat\Gamma_{(U+L)^P}$
($-13.5\%$) and smallest for $ \hat\Gamma_{U^P} $ ($-6.2\%$). When
normalized to the total rate, as is appropriate for density matrix elements,
the corresponding density matrix elements are reduced by $2.84\%$ and
increased by $5.38\%$ in magnitude by the radiative corrections, respectively.

The nonperturbative corrections range from $-0.9\%$ for the spin-dependent
corrections to a maximal $-3.4\%$ for the kinetic energy correction to
$\hat\Gamma_L$. The nonperturbative corrections are all negative except for
the kinetic energy correction to $\hat\Gamma_U$ and $\hat\Gamma_{I^P}$.

As specified in Eqs.~(\ref{singleangle-star}) and (\ref{singleangle}), the
asymmetry parameters $\alpha_P(D_s,D_s^*)$ can be measured in the
semi-inclusive decays of a polarized $\Lambda_b$ into the two decay channels.
For the pseudoscalar case the Born term level asymmetry $\alpha_P(D_s)=0.99$
is quite close to its maximal attainable value of 1 which would be achieved
for $y=0$. The Born term value is only slightly reduced to $\alpha_P(D_s)=0.97$
by the radiative and nonperturbative corrections. For the vector case the
asymmetry parameter is smaller. At Born term level one has
$\alpha_P(D_s^*)=0.37$ which is reduced to $\alpha_P(D_s^*)=0.35$ including
the radiative and nonperturbative corrections. As outlined in Sec.~2 the
transverse/longitudinal composition of the $D_s^*$ can be measured by the
$\cos\theta$-dependence in the angular decay distribution of its decay
products. At the Born term level the transverse/longitudinal composition is
given by $\hat\Gamma_U/\hat\Gamma_L=0.55$. This ratio is shifted upward by the
insignificant amount of $0.3\%$ through the radiative corrections. Adding all
corrections one finds a $7.3\%$ enhancement in the $U/L$ ratio.

Next we turn to our numerical results for the Cabibbo-suppressed
semi-inclusive decays $\Lam_b\rightarrow X_u+(D_s, D_s^*)$ induced by the
$b\rightarrow u$ transitions. Compared to the above Cabibbo-enhanced
semi-inclusive decays they are down by a factor
$(V_{ub}/V_{cb})^2\approx 10^{-2}$, which is only slightly compensated for by
a kinematical enhancement factor of $\approx 1.5$. Setting $m_u = 0$, i.e.\
$y=0$, one has
\begin{eqnarray}
  \hat\Gamma_{S^{\phantom P} \phantom{+ L}} & = & \phantom{+0.0000 \,}
  (1-0.1694-0.0130+0), \nonumber\\[2mm]
  \hat\Gamma_{U^{\phantom P} \phantom{+ L}} & = & \pp
  0.2750\,(1-0.1150+0.0285+0), \nonumber\\[2mm]
  \hat\Gamma_{L^{\phantom P} \phantom{+ L}} & = & \pp
  0.7250\,(1-0.1777-0.0267+0), \nonumber\\[2mm]
  \hat\Gamma_{U+L^{\phantom P}} & = & \phantom{+0.0000 \,}  
  (1-0.1605-0.0130+0), \nonumber\\[2mm]
  \hat\Gamma_{S^P \phantom{+ L}} & = & \phantom{+0.0000 \,}
  (1-0.1745-0.0212-0.0087), \\[2mm]
  \hat\Gamma_{U^P \phantom{+ L}} & = & -
  0.2750\,(1-0.1275-0.0212-0.0087), \nonumber\\[2mm]
  \hat\Gamma_{L^P \phantom{+ L}} & = & \pp
  0.7250\,(1-0.1800-0.0212-0.0087), \nonumber\\[2mm]
  \hat\Gamma_{(U+L)^P{\phantom P}} & = & \pp
  0.4500\,(1-0.2121-0.0212-0.0087), \nonumber\\[2mm]
  \hat\Gamma_{I^P \phantom{+ L}} & = & -
  0.2233\,(1-0.1506+0.0005-0.0087). \nonumber 
\end{eqnarray}
In the $b\rightarrow u$ case the radiative corrections and their spread are
larger than in the $b\rightarrow c$ case. The reduction of the partial rates
from the radiative corrections now scatter around $-17\%$, where the reduction
is largest for $\hat\Gamma_{(U+L)^P}$ ($-21.2\%$) and smallest for
$\hat\Gamma_U$ ($-11.5\%$). When normalized to the total rate, the
corresponding density matrix elements are reduced by $6.1\%$ and increased by
$5.4\%$ in magnitude by the radiative corrections, respectively. The dominance
of the longitudinal rate is now more pronounced. At the Born term level one
finds $\Gamma_U/\Gamma_L=2x^2=0.38$. The ratio $\Gamma_U/\Gamma_L$ is shifted
upward by $7.6\%$ by the radiative corrections. Adding up all corrections one
finds a $14.8\%$ upward shift for this ratio. For the asymmetry parameter one
obtains $\alpha_P(D_s)=0.984$ including all corrections which shifts the
uncorrected result $\alpha_P(D_s)=1$ downward by $1.6\%$. For the asymmetry
parameter $\alpha_P(D_s^*)$ one obtains $\alpha_P(D_s^*)=0.42$ which is lower
than the uncorrected result of $\alpha_P(D_s^*)=0.45$ by $7.3\%$. Let us
mention that our $O(\alpha_s)$ results on $\Gamma_{U+L}$ and $\Gamma_S$
numerically agree with the results of \cite{aleksan00} for both the
$b\rightarrow c$ and $b\rightarrow u$ transitions.

As a last point we want to discuss the semi-inclusive decays
$\Lam_b\rightarrow X_c+(\pi^-,\rho^-)$ which have not been discussed so far.
They are also induced by the diagrams Fig.~2 when the $c \rightarrow s$
transition in the upper leg is replaced by a $u \rightarrow d$ transition.
Using $f_{\pi^-}=132\mbox{ MeV}$, $f_{\rho^-}=216\mbox{ MeV}$ and
$V_{ud}=0.975$ one finds the Born term branching fractions
$\mbox{BR}_{b\rightarrow\pi^-+c}\cong 1.6\%$ and
$\mbox{BR}_{b\rightarrow\rho^-+c}\cong 4.6\%$.
In the latter case the rate is dominated by the longitudinal contribution
since $q^2=m_\rho^2$ is not far from $q^2=0$ where the rate would be entirely
longitudinal. In fact one finds $\Gamma_U/\Gamma_L=0.067$. It is important to
note that the diagrams Fig.~2 are not the only mechanisms that contribute to
the semi-inclusive decays $\Lam_b\rightarrow X_c+(\pi^-,\rho^-)$. Additional
$\pi^-$ and $\rho^-$ mesons can also be produced by fragmentation of the
$c$-quark at the lower leg.\footnote{As concerns the semi-inclusive decays
$\Lam_b\rightarrow X_c+(D_s^-,D_s^{\ast -})$ the possibility of producing
extra $D_s^-$ and $D_s^{*-}$ mesons through fragmentation of the $c$-quark is
ruled out for kinematic reasons.} As concerns the $\rho^-$ mesons resulting
from the fragmentation process they would not be polarized along their
direction of flight. This lack of polarization as compared to the strong
polarization of the $\rho$ mesons from the weak vertex could possibly be used
to separate $\rho^-$ mesons coming from the two respective sources.

\section{Summary and conclusions}

We have calculated the perturbative $O(\alpha_s)$ and the nonperturbative
$O(1/m_b^2)$ corrections to the seven structure functions that can be measured
in the semi-inclusive decay of a polarized $\Lam_b$ in the process 
$\Lam_b(\uparrow)\rightarrow X_c+(D_s^-,D_s^{*-})$. We have used the
factorization hypothesis to factorize the semi-inclusive decay into a
current-induced $\Lam_b\rightarrow X_c$ transition and a current-induced
vacuum one-meson transition. The dominant contribution to the current-induced
$\Lam_b\rightarrow X_c$ transition is given by the leading order HQET
transition $b\rightarrow c$. Thus the semi-inclusive decays of a polarized
$\Lam_b$ offer the unique opportunity to measure seven of the nine structure
functions that describe the current-induced free quark transition
$b\rightarrow c$.

We emphasize that there are also nonfactorizing $O(\alpha_s)$ contributions
which have not been included in our analysis. However, the nonfactorizing
$O(\alpha_s)$ contributions are colour suppressed and are thus expected to be
small.

We find that the perturbative corrections are always negative. The
nonperturbative corrections are negative in most of the cases. The net effect
of the corrections to the structure functions can become as large as $-20\%$
for the $b\rightarrow c$ transitions and can exceed $-20\%$ for the 
$b\rightarrow u$ transitions. When normalized to the total rate, as is
appropriate for density matrix elements accessible to experimental
measurement, the corrections become smaller but can still amount to
$\approx\pm 5\%$. 


\vspace{10mm}\noindent
{\bf Acknowledgements:} S.~Groote and M.C.~Mauser were supported by the DFG
(Germany) through the Graduiertenkolleg ``Eichtheorien'' at the University of
Mainz. 


\end{document}